# Broadband terahertz probes of anisotropic magnetoresistance disentangle extrinsic and intrinsic contributions

Lukáš Nadvorník[1,2,3,*], Martin Borchert[1,2,4], Liane Brandt[5], Richard Schlitz[6], Koen A. de Mare[7,8], Karel Výborný[7], Ingrid Mertig[5], Gerhard Jakob[9], Matthias Kläui[9], Sebastian T.B. Goennenwein[6], Martin Wolf[2], Georg Woltersdorf[5] and Tobias Kampfrath[1,2]

1. Department of Physics, Freie Universität Berlin, 14195 Berlin, Germany
2. Department of Physical Chemistry, Fritz Haber Institute of the Max Planck Society, 14195 Berlin, Germany
3. Faculty of Mathematics and Physics, Charles University, 121 16 Prague, Czech Republic
4. Max Born Institute for Nonlinear Optics and Short Pulse Spectroscopy, 12489 Berlin, Germany
5. Institut für Physik, Martin-Luther-Universität, Halle, Germany
6. Institut für Festkörper- und Materialphysik, Technische Universität Dresden, 01062 Dresden, Germany
7. Institute of Physics, Academy of Sciences of the Czech Republic, v.v.i., 162 00 Prague, Czech Republic
8. Department of Applied Physics, Eindhoven University of Technology, Eindhoven 5612 AZ, The Netherlands
9. Institut für Physik, Johannes Gutenberg-Universität Mainz, 55128 Mainz, Germany

[*] E-mail: nadvornik@karlov.mff.cuni.cz

Anisotropic magnetoresistance (AMR) is a ubiquitous and versatile probe of magnetic order in contemporary spintronics research. Its origins are usually ascribed to extrinsic effects (i.e. spin-dependent electron scattering), whereas intrinsic (i.e. scattering-independent) contributions are neglected. Here, we measure AMR of polycrystalline thin films of the standard ferromagnets Co, Ni, $Ni_{81}Fe_{19}$ and $Ni_{50}Fe_{50}$ over the frequency range from DC to 28 THz. The large bandwidth covers the regimes of both diffusive and ballistic intraband electron transport and, thus, allows us to separate extrinsic and intrinsic AMR components. Analysis of the THz response based on Boltzmann transport theory reveals that the AMR of the Ni, $Ni_{81}Fe_{19}$ and $Ni_{50}Fe_{50}$ samples is of predominantly extrinsic nature. However, the Co thin film exhibits a sizeable intrinsic AMR contribution, which is constant up to 28 THz and amounts to more than 2/3 of the DC AMR contrast of 1%. These features are attributed to the hexagonal structure of the Co crystallites. They are interesting for applications in terahertz spintronics and terahertz photonics. Our results show that broadband terahertz electromagnetic pulses provide new and contact-free insights into magneto-transport phenomena of standard magnetic thin films on ultrafast time scales.

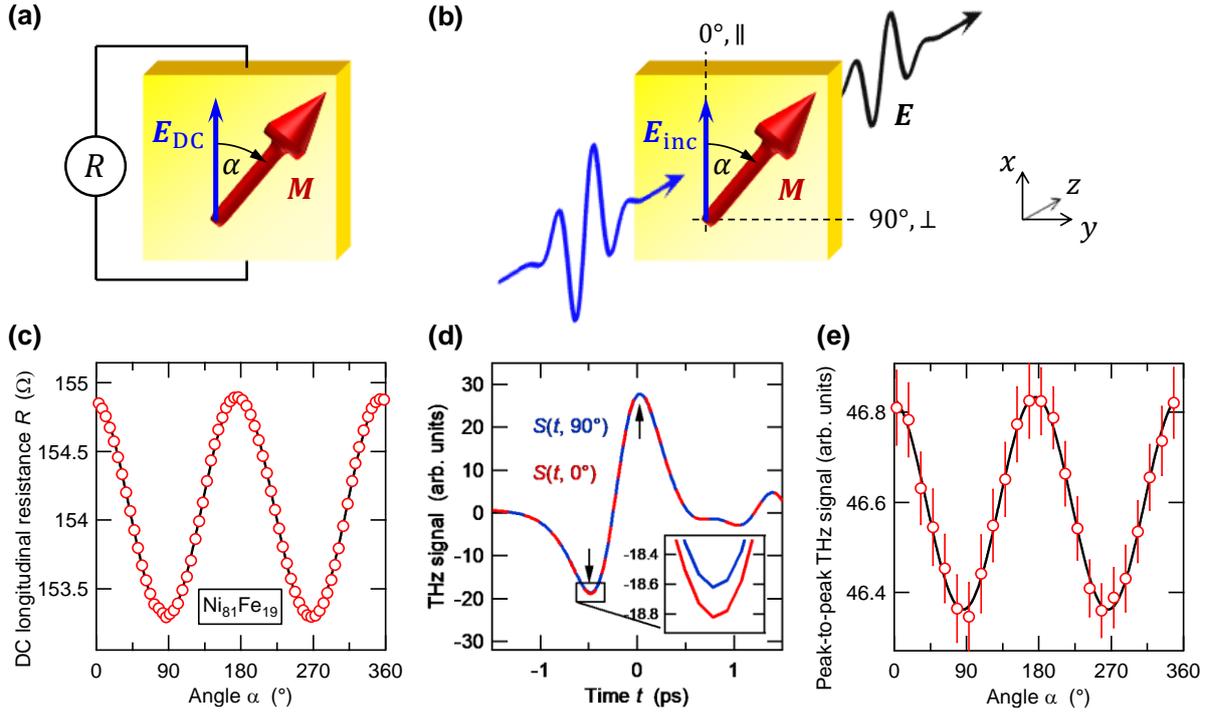

FIG. 1. Measuring DC and THz AMR. (a) Schematic of a DC electrical AMR measurement. The resistance $R$ of a magnetic thin film with magnetization $\mathbf{M}$ (red arrow) along the applied DC electric field $\mathbf{E}_{\mathrm{DC}}$ (blue arrow) is measured for different rotation angles $\alpha$ of $\mathbf{M}$. (b) Schematic of the THz AMR measurement. An $x$-polarized THz pulse with transient electric field $\mathbf{E}_{\mathrm{inc}}(t)$ (blue arrow) is incident on the magnetic thin film. After traversal of the sample, we detect the $x$ component of the THz electric field $\mathbf{E}$, that is, $\mathbf{E}$ projected onto the fixed direction of $\mathbf{E}_{\mathrm{inc}}$, as a function of the magnetization angle $\alpha$. (c) DC longitudinal resistance of $\mathrm{Ni}_{81}\mathrm{Fe}_{19}$ vs $\alpha$ (red open circles) with fit by $R_\perp + \Delta R \cos^2 \alpha$ (black solid line). (d) THz waveforms $S(t, \alpha)$ for $\alpha = 90°$ (blue solid line) and $0°$ (red dashed line) vs time $t$ for the emitter-detector configuration covering the frequency range 0.2…2THz. The inset shows a magnified version of the signal around its minimum, indicating a signal change of about 1%. (e) Peak-to-peak amplitude of the THz signal waveform $S(t, \alpha)$ as a function of $\alpha$ (open circles) with fit as in panel (c). The signal extrema used for the evaluation are indicated by the black arrows in panel (d).

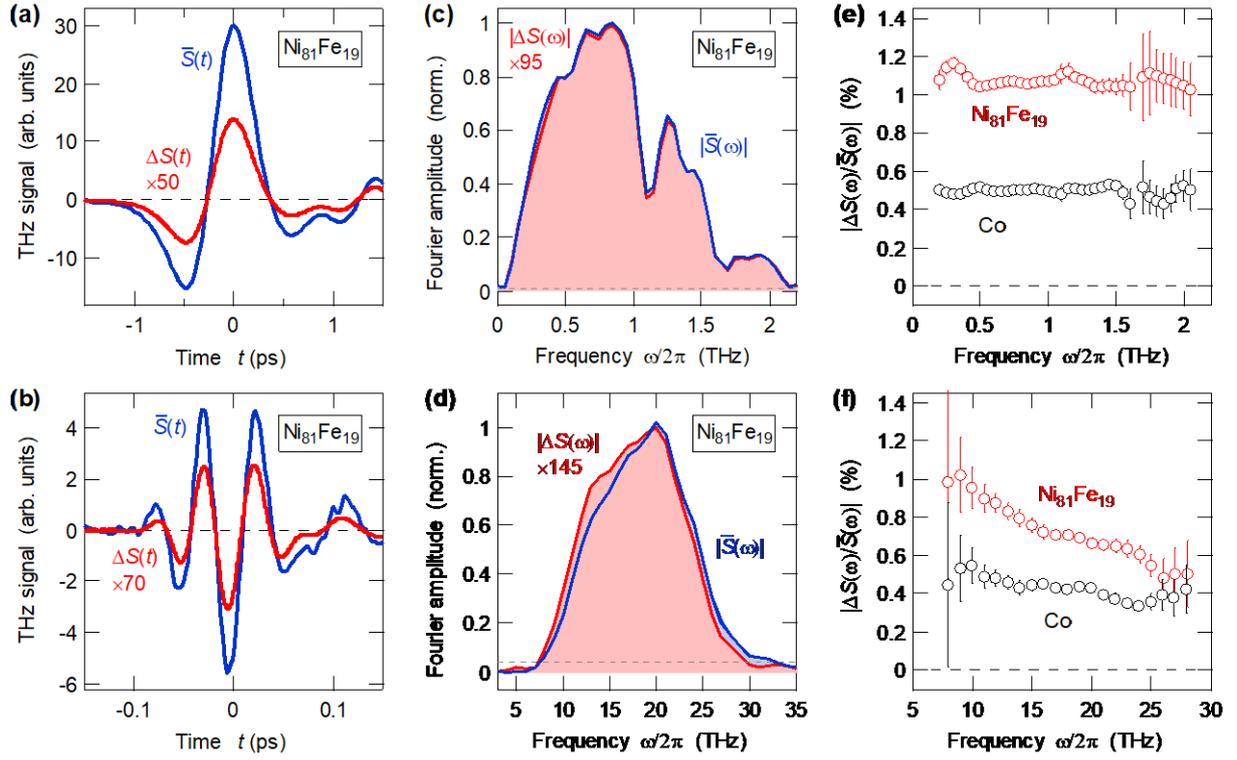

FIG. 2. THz AMR probes of $Ni_{81}Fe_{19}$ and Co. (a) THz signal waveforms $\Delta S(t) = S(t, 0°) - S(t, 90°)$ (red line) and $\bar{S}(t) = [S(t, 0°) + S(t, 90°)]/2$ (blue line) of a THz pulse from a source-detector combination with a bandwidth 0.2…2 THz after having traversed the $Ni_{81}Fe_{19}$ thin film. The difference waveform $\Delta S(t)$ reports on the change in the sample transmission when the magnetization is rotated from $\alpha = 90°$ to $0°$ [see Fig. 1(b)]. It is a signature of the AMR. (b) Same as panel (a), but for measurements with bandwidth 8…28 THz. (c,d) Fourier amplitude spectra $|\Delta S(\omega)|$ and $|\bar{S}(\omega)|$ of the traces of panel (a,b), respectively. Signals are normalized to their respective maximum. The mean noise level of $|\Delta S(\omega)|$ is indicated by the grey dashed line. (e,f) Relative spectral amplitude changes $|\Delta S(\omega)/\bar{S}(\omega)|$ as derived from panel (c,d), respectively (red circles). Results for Co are also shown (black circles). Note that $|\Delta S(\omega)/\bar{S}(\omega)|$ is closely related to the AMR contrast through Eq. (6). The error bars indicate the precision of the measurements as estimated from the noise level in panels (c,d). Data in all panels are scaled by the indicated factors for clarity.

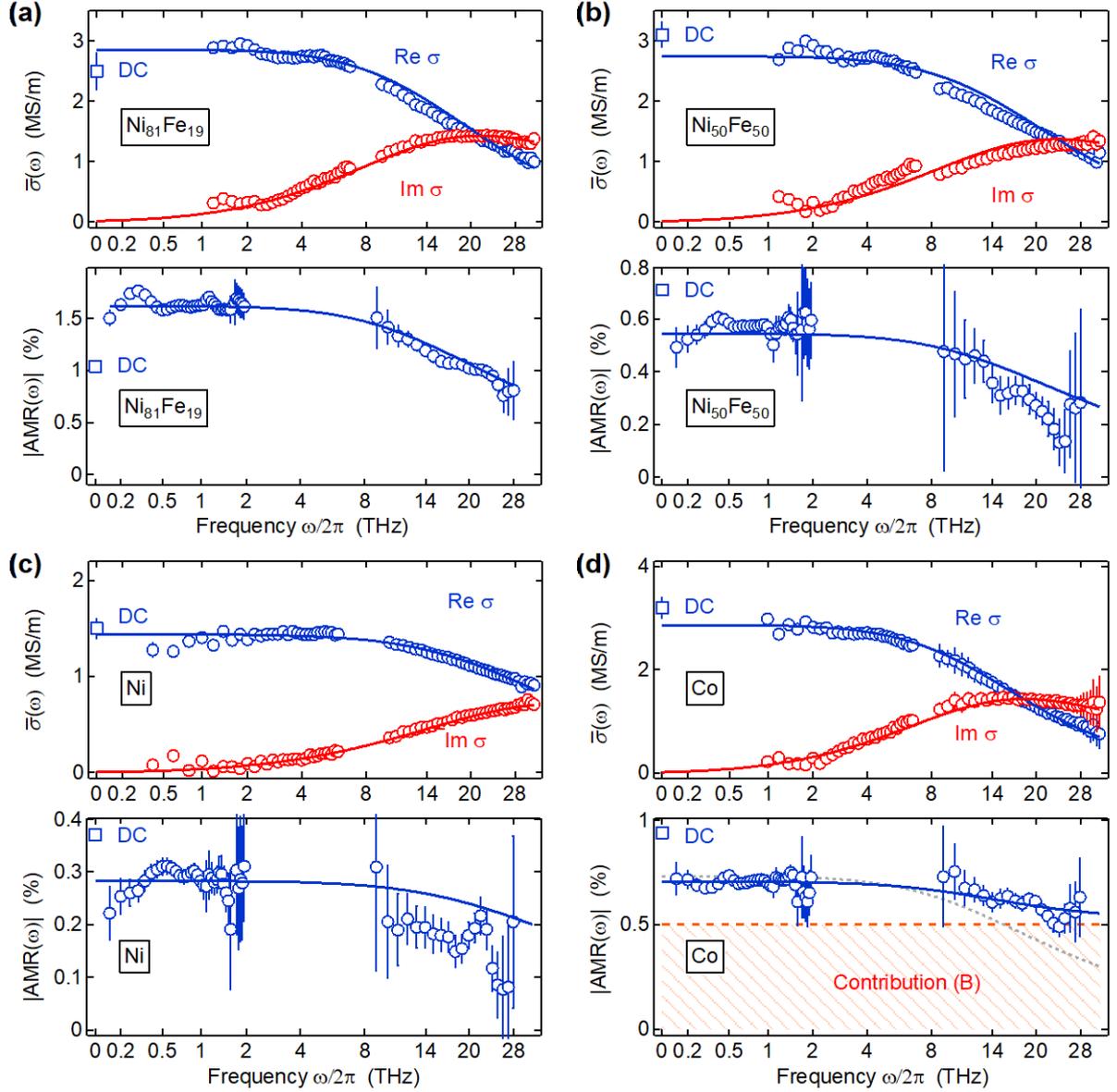

FIG. 3. Mean conductivity and AMR contrast from DC to 28 THz. (a) Mean complex-valued conductivity $\bar{\sigma} = (\sigma_\parallel + \sigma_\perp)/2$ (upper panel) and magnitude |AMR| of the AMR contrast (lower panel) of the $Ni_{81}Fe_{19}$ thin film as a function of frequency. Solid lines are fits based on the Drude formula [Eqs. (2) and (3)]. DC values are shown at the origin of the frequency axis (square symbol). (b,c,d) Same as panel (a), but for $Ni_{50}Fe_{50}$, Ni and Co, respectively. In panel (d), the red dashed line and shaded area indicate the intrinsic contribution (B) with amplitude $B = 0.5\ \%$ to AMR. For comparison, the grey dotted curve shows a fit using Eq. (3) without intrinsic contribution ($B = 0$). For clarity, the scaling of the frequency axis is nonlinear ($\propto \omega^{-1/4}$). The measured frequency ranges are determined by the spectral coverage of the THz emitter-detector configurations used (see Section III.B). The error bars of the THz data indicate their precision as estimated from repeated measurements (for $\bar{\sigma}$) and the spectral noise floor [for |AMR|, see Figs. 2(c,d)]. The overall scaling of |AMR| is accurate within an estimated uncertainty of 30% (see Appendix A).

|  | **Ni$_{81}$Fe$_{19}$** | **Ni$_{50}$Fe$_{50}$** | **Ni** | **Co** |
|---|---|---|---|---|
| Crystallite symmetry | Cubic | Cubic | Cubic fcc | Hexagonal hcp |
| $\sigma_{\mathrm{DC}}$ (MS/m) | 2.5±0.3 | 3.1±0.2 | 1.5±0.1 | 3.2±0.2 |
| $\bar{\sigma}(0)$ (MS/m) | 2.9±0.2 | 2.8±0.1 | 1.4±0.2 | 2.9±0.2 |
| $\bar{\tau}$ (fs) | 9±2 | 9±2 | 5±1 | 11±2 |
| $\Delta\sigma_{\mathrm{DC}}/\sigma_{\mathrm{DC}}$ (%) | 1.0±0.1 | 0.7±0.1 | 0.37±0.05 | 0.9±0.1 |
| $\Delta\sigma(0)/\bar{\sigma}(0)$ (%) | 1.6±0.4 | 0.5±0.1 | 0.3±0.1 | 0.7±0.2 |
| $B = -\Delta(\Omega_{\mathrm{pl}}^2)/\bar{\Omega}_{\mathrm{pl}}^2$ | (0.0±0.1)×10$^{-2}$ | (0.0±0.1)×10$^{-2}$ | (0.00±0.07)×10$^{-2}$ | (0.5±0.1)×10$^{-2}$ |
| $A = -\Delta\tau/\bar{\tau}$ | (1.6±0.2)×10$^{-2}$ | (0.5±0.1)×10$^{-2}$ | (0.28±0.07)×10$^{-2}$ | (0.2±0.1)×10$^{-2}$ |
| $B/A$ | 0.0±0.1 | 0.0±0.2 | 0.0±0.3 | 2.5±1.5 |

TABLE I. Sample properties and fit parameters. All films are polycrystalline. The parameters $\bar{\sigma}(\omega = 0)$, $\bar{\tau}$, $\Delta\sigma(0)/\bar{\sigma}(0)$, $A$ and $B$ were obtained by fitting the Drude model for conductivity [Eq. (2)] and AMR contrast [Eq. (3)] to the data shown in Fig. 3. The ratio $B/A$ of extrinsic and intrinsic contributions to the AMR reveals a different AMR regime in Co as compared to the other materials. The parameters $A$ and $B$ were obtained by fitting over the whole frequency range 0.2…28 THz and over the range 8…28 THz, both yielding consistent values (see Supplemental Note 1). The uncertainties arise from the precision and accuracy of our measurements as detailed in Appendix A.

## I. INTRODUCTION

The electrical resistance of a ferromagnet along the applied electric field is known to depend on the direction of the magnetization *M* [Fig. 1(a)]. This anisotropic magnetoresistance (AMR)[1,2,3,4,5,6,7] is a well-studied magnetoresistive effect and a powerful tool to detect the magnetic order parameter of ferromagnets as well as ferrimagnets[8,9]. As AMR is even in the magnetic order parameter, it has lately received additional attraction as a probe of the Néel vector of antiferromagnets[10]. Therefore, AMR has large potential for applications in future spintronic devices[11].

The canonic way to describe the origins of AMR relies on an extrinsic mechanism, that is, spin-dependent electron scattering due to crystal imperfections such as impurities and phonons. In transition metals, the *M*-dependent rate of electron scattering out of the current-carrying s-states is understood to arise from spin-orbit coupling, which reduces the symmetry of the target d-states[2,3,5,12]. Other extrinsic scenarios involve magnetic impurities acting on spin-orbit-coupled p-states (in, for instance, dilute magnetic semiconductors[13]) and non-magnetic impurities acting on states with isotropic band dispersion but anisotropic wavefunctions[14].

Only recently, theoretical works pointed out that AMR can already be significant in perfect crystals. An example of such intrinsic (i.e. scattering-independent) mechanism is a change in the group velocity of Bloch states due to spin-orbit coupling.[15,16,17,18,19,20]. First signatures of intrinsic contributions to DC AMR were reported[21,22] based on extensive electric transport measurements and *ab initio* theory. These highly promising results also show that more direct and versatile experimental methods are required to extract extrinsic and intrinsic AMR contributions.

To straightforwardly separate extrinsic (scattering-dependent) and intrinsic (scattering-independent) electron transport, we propose to probe the AMR dynamics on time scales both slower and faster than the time scale $\tau$ at which electron scattering takes place. To implement this idea, AMR needs to be measured over a wide frequency range from DC to several 10 THz. The lower frequencies $\omega/2\pi$ of this interval probe diffusive (i.e. scattering-dominated) transport in which an electron undergoes many collisions during one oscillation of the probing electric field ($\omega\tau \ll 1$). In contrast, the frequencies at the higher end are more sensitive to electron motion in the ballistic limit (i.e. without scattering) because the probing electric field oscillates many times between subsequent electron collisions ($\omega\tau \gg 1$). So far, measurements of AMR of common magnets were reported at either DC, at frequencies around 1 THz (Refs. 23, 24, 25) or in the infrared, where AMR is usually referred to as magnetic linear birefringence[26,27].

In this work, we measure AMR of common ferromagnets in the regime of both ballistic and diffusive electron transport by means of low-noise broadband THz spectroscopy from DC to 28 THz. A frequency-resolved data analysis based on Boltzmann transport theory allows us to robustly separate the (A) extrinsic and (B) intrinsic components of AMR. We find that component (B) is significant and even dominates the total AMR for the case of polycrystalline Co. Using numerical estimates, we attribute this observation to the hexagonal structure of the Co crystallites. Owing to its instantaneous response up to at least 28 THz, the intrinsic AMR of Co is highly interesting for applications in future THz spintronic devices. Our results also highlight that broadband THz AMR is a powerful and versatile probe of ultrafast spin dynamics.

## II. AMR IN THE DRUDE-BOLTZMANN FRAMEWORK

The AMR contrast at frequency $\omega/2\pi$ is defined as

$$\text{AMR}(\omega) = -\frac{\sigma_\parallel - \sigma_\perp}{\sigma_\perp} \approx -\frac{\Delta\sigma}{\bar{\sigma}} \qquad (1)$$

where $\sigma_j(\omega)$ is the conductivity for the magnetization $\mathbf{M}$ parallel ($j = \parallel$) or perpendicular ($j = \perp$) to the applied electric field amplitude $\mathbf{E}(\omega)$. Typically, the DC AMR contrast ($\omega = 0$) is positive and reaches values of the order of 1% to 10% (Ref. 2). Therefore, the difference $\Delta\sigma = \sigma_\parallel - \sigma_\perp$ is relatively small, and $\sigma_\parallel$ and $\sigma_\perp$ are very close to the mean conductivity $\bar{\sigma} = (\sigma_\parallel + \sigma_\perp)/2$. We note that the $\sigma_j(\omega)$, like all frequency-domain quantities, are generally complex-valued.

In contrast to the anomalous Hall effect (which is of first order in $\mathbf{M}$), there are significantly less theoretical studies of the microscopic mechanism of AMR (which is quadratic in the order parameter). A frequently used theoretical approach is based on the Boltzmann equation describing intraband transport[14,17,21,28,29]. Assuming state-independent relaxation rates, one can derive the Drude formula[30,31,32,,33]

$$\sigma_j(\omega) = \frac{\sigma_j(0)}{1 - i\omega\tau_j} = \frac{1}{Z_0 c} \frac{\Omega_{\text{pl}j}^2}{\tau_j^{-1} - i\omega} \tag{2}$$

where $j = \parallel$ or $\perp$, $\omega/2\pi$ is the frequency of the driving field, $\sigma_j(0)$ equals the DC conductivity, and $\tau_j$ is the current relaxation time.

The second part of Eq. (2) is a rewritten Drude formula where $Z_0 \approx 377\,\Omega$ is the free-space impedance, $c$ is the speed of light, and $\Omega_{\text{pl}j}/2\pi$ is the plasma frequency. This formulation allows us to identify (A) extrinsic contributions (due to electron scattering) and (B) intrinsic (scattering-independent) contributions to the AMR contrast. Inequality $\sigma_\parallel \neq \sigma_\perp$ of the $\parallel$ and $\perp$ conductivities and, thus, AMR can arise from the $\mathbf{M}$-direction dependence of (A) the current relaxation time ($\tau_\parallel \neq \tau_\perp$) and (B) the plasma frequency ($\Omega_{\text{pl}\parallel} \neq \Omega_{\text{pl}\perp}$). Because $\Omega_{\text{pl}j}^2$ is given by a summation of the squared electron band velocity component $j$ over the Fermi surface[34,35,36,37], it is a measure of the weight of intrinsic (scattering-independent) contributions to the conductivity. In contrast to $\Omega_{\text{pl}j}^2$, the velocity relaxation rate $\tau_j^{-1}$ arises from electron-impurity and electron-phonon collisions and, thus, captures extrinsic (scattering-related) effects. Note that previous studies using the Boltzmann approach ascribed AMR to contribution (A)[2,28].

We expect that in the diffusive transport regime ($\omega\tau_j \ll 1$), both extrinsic and intrinsic effects contribute to AMR, whereas in the ballistic regime ($\omega\tau_j \gg 1$), intrinsic contributions should dominate. To put this expectation onto a quantitative basis, we substitute the Drude formula [Eq. (2)] into the definition of the AMR contrast [see Eq. (1) and Appendix B]. The resulting relationship

$$\text{AMR} = \frac{A}{1 - i\omega\tau_\perp} + B \tag{3}$$

has remarkable implications: First, the two terms on its right-hand side scale with $A = -\Delta\tau/\tau_\perp$ and $B = -\Delta(\Omega_{\text{pl}}^2)/\Omega_{\text{pl}\perp}^2$ where $\Delta$ always refers to the difference of the $\parallel$ and $\perp$ component, for instance $\Delta\tau = \tau_\parallel - \tau_\perp$. Therefore, the $A$ and $B$ terms, respectively, quantify the (A) extrinsic (scattering-based) and (B) intrinsic (scattering-independent) contribution to AMR.

Second, as expected, the components (A) and (B) exhibit a distinctly different frequency dependence. The extrinsic contribution (A) rolls off with frequency just as the conductivity $\sigma_\perp$ [Eq. (2)] does. The frequency scale of this decrease is set by the velocity relaxation rate $\tau_\perp^{-1}/2\pi$, which is typically of the order of 10 THz[38,39]. The intrinsic contribution (B), in contrast, is $\omega$-independent, thereby making it interesting for potential high-frequency applications in THz spintronics.

Finally, Eq. (3) guides us how to determine the weight of the two AMR contributions: We need to conduct a sufficiently broadband AMR measurement. Our goal is, therefore, to measure the anisotropic conductivity of common ferromagnets over the broad range from $\omega/2\pi \sim 0$ to tens of THz.

### III. EXPERIMENTAL SETUP

#### A. Samples

As samples, we chose thin films of common ferromagnetic metals with in-plane magnetic anisotropy: $Ni_{81}Fe_{19}$ (thickness of 8 nm), $Ni_{50}Fe_{50}$ (10 nm), Ni (10 nm) and Co (10 nm). As detailed in Appendix A and Supplemental Note 3, they were grown by sputtering on isotropic Si substrates. Subsequently, the samples were cut in two pieces to convey the DC and THz experiments. For the THz experiments, a part of the substrate was not covered by the metal layer to permit reference transmission measurements to extract the THz conductivity.

All thin films were prepared in the polycrystalline phase. As compiled by Table 1, they consist of crystallites having cubic ($Ni_{81}Fe_{19}$, $Ni_{50}Fe_{50}$, Ni) or hexagonal symmetry (Co). Because the size of the randomly oriented crystals is orders of magnitude smaller than the wavelength of the probing THz radiation, all films are macroscopically isotropic in the sample plane in the absence of magnetic order ($\boldsymbol{M} = 0$). A symmetry analysis of our samples (see Appendix C) shows that the in-plane conductivity tensor is fully determined by the two conductivity values $\sigma_\parallel$ and $\sigma_\perp$ parallel and perpendicular to the magnetization, independent of the sample azimuth. The difference $\sigma_\parallel - \sigma_\perp$ equals $2\langle G \rangle_{xyxy} \boldsymbol{M}^2$ where $\langle G \rangle_{xyxy}$ is the only relevant element of the rotationally averaged AMR tensor.

#### B. Conductivity and AMR measurements

**DC measurements.** In general, determination of the AMR contrast [Eq. (1)] of our samples relies on measuring the ratio of the conductivities for $\boldsymbol{M} \parallel \boldsymbol{E}$ and $\boldsymbol{M} \perp \boldsymbol{E}$. At DC frequency, this goal was achieved by a four-point approach[40]. A rectangular piece was cleaved from the sample and contacted in the corners. A constant current was applied along the longer side, and the voltage drop and, thus, resistance $R(\alpha)$ along this fixed direction was measured as a function of the angle $\alpha$ of the in-plane magnetization $\boldsymbol{M}$ [see Fig. 1(a)]. The AMR contrast [Eq. (1)] is given by $-\Delta\sigma_{DC}/\overline{\sigma}_{DC} = \Delta R/\overline{R}$ where $\Delta R$ and $\overline{R}$ are, respectively, the modulation depth and mean value of $R(\alpha)$.

Note that the mean DC conductivity $\overline{\sigma}_{DC}$ of the metal film is related to $\overline{R}$ through an unknown factor that is given by the current distribution. We, consequently, used the van-der-Pauw method[41,42] to measure $\overline{\sigma}_{DC}$.

**THz measurements.** To determine the AMR contrast of our samples at THz frequencies, we did not use any electrical contacts and measured the transmission of a broadband THz electromagnetic pulse through the specimen in a quasi-optical manner [Fig. (1b)]. To this end, THz pulses were obtained by difference-frequency generation of femtosecond laser pulses (duration of 10 fs, center wavelength of 800 nm, energy of 1 nJ) from a Ti:sapphire laser oscillator (repetition rate of 80 MHz) in a suitable nonlinear-optical material. The THz pulses were linearly polarized along the $x$ axis and normally incident onto the sample [Fig. 1(b)].

After transmission through the sample, a wire-grid polarizer projected the THz field $\boldsymbol{E}$ onto the $x$ axis, that is, the polarization direction of the incident THz electric field. The THz pulses were detected by electrooptic sampling using a suitable electrooptic crystal[43]. The resulting THz signal $S$ vs time $t$ is related to the THz electric field component $E_x(t)$ directly behind the sample [Fig. 1(d)] by a linear transfer function that cancels in the subsequent data analysis.

To ensure optimum frequency coverage and signal-to-noise ratio, we used various combinations of THz sources and detectors. For AMR measurements, we used a bias-free bimetallic emitter (TeraBlast, Protemics GmbH) and 1 mm thick ZnTe(110) crystal as detector for the range 0.2….2 THz, while a

90 µm thick GaSe emitter and a 10 µm thin ZnTe detection crystal was employed for the range 8…28 THz. This combination delivers sufficient THz signal amplitude to resolve the small AMR-induced changes of the sample transmission upon rotation of the magnetization from 0° to 90°.

For measurement of the mean (diagonal) conductivity $\bar{\sigma}$, where signal amplitudes are sufficiently large, we replaced the bimetallic emitter by a spintronic THz emitter[43] (TeraSpinTec GmbH) and used a 250 µm thick GaP crystal as detector. This combination delivers an order of magnitude less signal amplitude but covers the range 1…6 THz, which is useful for the precise determination of parameters of the Drude formula [Eq. (2)].

Typical examples of transmitted THz signal waveforms are shown in Fig. 1(d) as well as in Figs. 2(a) and 2(b) (blue curves). As detailed in Section IV and Appendix A, the measured THz transmission signals can be used to determine the mean THz conductivity $\bar{\sigma}$ of a thin metallic layer. Similarly, by modulating the magnetization angle $\alpha$ between 90° and 0°, we can infer the THz AMR contrast.

### C. Magnetization control

**Slow modulation.** The magnetization angle $\alpha$ relative to the fixed direction of the applied DC or THz electric field [see Figs. 1(a) and 1(b)] was controlled by a suitable external magnetic field. For the DC measurements, we used a magnetic field of 1.1 T from a Halbach array of permanent magnets that was slowly rotated about the sample. For the THz measurements as a function of all magnetization angles $\alpha$ between 0 and 360°, we employed a rotatable pair of permanent magnets with a field of approximately 40 mT at the sample position.

**Fast modulation.** To drastically enhance the signal-to-noise ratio of the THz AMR measurements, we modulated the magnetization angle $\alpha$ at kilohertz rates by superimposing a sinusoidal AC magnetic field (frequency of 6 kHz) from an electromagnet and a perpendicular DC magnetic field from a permanent magnet. As the two fields had an amplitude of approximately 30 mT at the sample position, the magnetization angle $\alpha$ was varied between $\alpha_{\min} \approx 0°$ and $\alpha_{\max} \approx 90°$, that is, between approximately parallel and perpendicular to the polarization of the THz wave [see Fig. 1(b)]. Lock-in-type phase-sensitive demodulation of the THz signal allowed us to extract its magnetic-field- and, thus, AMR-induced signal variations.

The magnetic field strength of the 6 kHz arrangement was sufficient to fully saturate the sample magnetization as confirmed by measuring the magnetization direction by THz emission spectroscopy[43]. Note that the expected AMR signal is determined by the sample magnetization rather than the external magnetic field, which induces only isotropic[2,44,45] and, thus, negligible, conductivity changes. Therefore, our various methods of magnetization modulation deliver conductivity modulations that can directly be compared to each other.

### IV. RESULTS

### A. Impact of magnetization direction

To study the sample conductivity as a function of the magnetization angle $\alpha$ [Fig. 1(b)], we varied the direction of the external magnetic field by the slowly rotating permanent magnets. Figure 1(c) shows the measured DC resistance of the $Ni_{81}Fe_{19}$ thin film vs $\alpha$. We observe the typical $\cos^2 \alpha$-like resistance modulation that is expected for samples described by two conductivities $\sigma_\parallel$ and $\sigma_\perp$ (Ref. 2). Indeed, a fit by $R_\perp + \Delta R \cos^2 \alpha$ yields excellent agreement with the experimental data. From the modulation depth and the average resistance $\bar{R} = [R(0°) + R(90°)]/2$, we estimate an AMR contrast $\Delta R/\bar{R} \approx -\Delta\sigma_{\mathrm{DC}}/\bar{\sigma}_{\mathrm{DC}}$ of approximately 1%. The DC AMR data for the other samples are shown in Supplemental Figure S1 while the fit parameters are displayed in Table 1.

We now turn to the THz measurements. Figure 1(d) displays the signals $S(t, \alpha)$ of THz waveforms after traversal of the sample for $\alpha = 0°$ and $90°$. For these measurements, the emitter-detector pair covering the range 0.2…2 THz was used. While the two signals are nearly identical, a magnified plot around the signal minimum reveals that the signal for $\alpha = 0°$ has larger amplitude than the signal for $\alpha = 90°$. This observation is consistent with the DC measurements [Fig. 1(c)] and Eq. (A2): Changing the magnetization angle from $\alpha = 0°$ to $90°$ yields a smaller sample resistance and, thus, larger conductivity, resulting in better screening of the incident THz field and, therefore, in a smaller THz field amplitude behind the sample.

To complete the picture, we determined the peak-to-peak amplitude of all THz signals $S(t, \alpha)$ as indicated by the two black arrows in Fig. 1(d). The resulting THz peak-to-peak amplitude is displayed in Fig. 1(e) as a function of the magnetization angle $\alpha$. It exhibits the same $\alpha$ dependence and comparable contrast as the DC resistance [Fig. 1(c)]. Again, a $\cos^2 \alpha$ fit yields excellent agreement with the experimental data [Fig. 1(e)]. We explicitly confirmed that the $\alpha$-dependent signal component disappeared when either (i) test samples without magnetic layer were used, (ii) the strength of the magnetic field was lowered below a critical value or (iii) the THz beam was blocked.

We conclude that the $\alpha$-dependent THz signal arises from the anisotropic conductivity of the magnetic thin film under study. As the $\alpha$-dependence and relative magnitude of this signal [Fig. 1(e)] coincide with that of the DC AMR signal [Fig. 1(c)], we assign the $\alpha$-dependent THz signal modulation to the AMR effect at THz frequencies.

## B. THz AMR differential spectra

To enable spectral analysis of the THz AMR with strongly increased signal-to-noise ratio, we modulated the magnetization angle $\alpha$ at a frequency of 6 kHz between $\alpha_{\min} \approx 0°$ and $\alpha_{\max} \approx 90°$ [see Fig. 1(b)]. By demodulation with a lock-in-type technique, we obtain the difference signal

$$\Delta S(t) = S(t, \alpha_{\min}) - S(t, \alpha_{\max}), \qquad (4)$$

while in a separate measurement, the mean signal

$$\bar{S}(t) = \frac{S(t, \alpha_{\min}) + S(t, \alpha_{\max})}{2} \qquad (5)$$

is acquired.

Typical time-domain raw data and their spectra are shown in Figs. 2(a)-2(d) for the case of the $Ni_{81}Fe_{19}$ thin film. While Fig. 2(a) displays the signals $\Delta S(t)$ and $\bar{S}(t)$ for an incident 0.2…2 THz pulse, Fig. 2(b) shows corresponding traces for a 8…28 THz pulse. By Fourier-transformation of the data of Figs. 2(a) and 2(b), the amplitude spectra $\Delta S(\omega)$ and $\bar{S}(\omega)$ of Figs. 2(c) and 2(d) are obtained. Relative $\alpha$-induced spectral amplitude changes $|\Delta S(\omega)/\bar{S}(\omega)|$ are displayed in Figs. 2(e) and 2(f) for $Ni_{81}Fe_{19}$ and Co, respectively.

The raw data of Figs. 2(c) and 2(d) reveal an interesting behavior: Both $\Delta S(\omega)$ and $\bar{S}(\omega)$ have exactly the same spectrum at 0.2…2 THz [Fig. 2(c)], but differ noticeably at 8…28 THz [Fig. 2(d)]. Figures 2(e) and 2(f) confirm this observation: $|\Delta S(\omega)/\bar{S}(\omega)|$, which scales with the AMR magnitude [see Eq. (6)], is independent of frequency below 2 THz [Fig. 2(e)], but starts decreasing above 2 THz, eventually reducing to about 50% at 20 THz [Fig. 2(f)]. This behavior is consistent with the amplitudes of the time-domain data [Figs. 2(a) and 2(b)] and their spectra [Figs. 2(c) and 2(d)]. Thus, Figs. 2(e) and 2(f) show that AMR is operative at frequencies up to 30 THz, but decreases on a scale of ~10 THz which coincides with typical current relaxation rates[38,39].

## C. From signals to conductivities and AMR

**Mean conductivities.** To better understand these observations, we also determined the mean conductivity $\bar{\sigma}$ of our samples at 1...6 THz and 8...28 THz (see Section III.B). For this purpose, we measured the signals $\bar{S}(t)$ [mean signal of Eq. (5) with respect to the full sample] and the reference signal $S_{\text{ref}}(t)$ corresponding to transmission through the plain substrate in sample regions without metal film. Using the Tinkham formula[46] (see Appendix A), we obtained the mean conductivity of the metal layer.

Real and imaginary part of the mean conductivity $\bar{\sigma}$ vs frequency $\omega/2\pi$ are displayed in Figs. 3(a-d) (top panels) along with the DC conductivity $\sigma_{\text{DC}}$ for all four samples studied. We note that the measured DC conductivity agrees well with the THz mean conductivity between 1 and 4 THz.

To gain access to microscopic parameters, we fit the measured conductivities using the Drude formula [Eq. (2)]. As shown by the solid lines of Figs. 3(a)-3(d), the Drude-Boltzmann framework provides a very good description of our experimental data over more than two frequency decades. Broadband Drude-like behavior of metals is quite common and was previously observed also for other magnetic thin films[47,48,49,50], magnetic multilayers[33] and nonmagnetic metals[30,51]. The best-fit parameters of our data, the mean zero-frequency conductivity $\bar{\sigma}(0)$ and the mean scattering rate $\bar{\tau}$, are summarized in Table 1. Again, we obtain a good match between the measured DC conductivity $\sigma_{\text{DC}}$ and the zero-frequency extrapolation $\bar{\sigma}(0)$. The current relaxation times $\bar{\tau}$ are found to be of the order of 10 fs, which is a typical value for metal thin films[33,38,39].

**AMR contrast.** To infer the AMR contrast [Eq. (1)], we use the $\bar{\sigma}(\omega)$ as determined by the fits above and the relationship (see Appendix A)

$$\text{AMR}(\omega) = \frac{\Delta S(\omega)}{\bar{S}(\omega)} \left[ 1 + \frac{n_S(\omega) + n_A(\omega)}{Z_0 d \bar{\sigma}(\omega)} \right]. \tag{6}$$

Here, $n_S$ and $n_A$ are the frequency-dependent refractive indices of air and substrate, and $d$ is the thickness of the metal layer.

The modulus $|\text{AMR}(\omega)|$ is displayed in Fig. 3 (bottom panels) vs frequency for all materials investigated. We see that the AMR contrast is approximately frequency-independent for $\omega/2\pi <$ 2 THz with magnitudes ranging from 0.3% (Ni) up to 1.6% (Ni$_{81}$Fe$_{19}$). These values are compatible with the DC quantities obtained by contact-based measurements within the uncertainties of our methodology. The various error sources are discussed in Appendix A.

We did not attempt to determine the phase of the AMR contrast because the signals $\Delta S$ and $\bar{S}$ were taken at different times. Therefore, the complex-valued ratio $\Delta S(\omega)/\bar{S}(\omega)$ may be subject to an unknown phase shift which does not allow us to determine the phase of $\text{AMR}(\omega)$ through Eq. (6). We emphasize that this lack of information is, however, no issue because the modulus of $\text{AMR}(\omega)$ is fully sufficient to determine the ratio $B/A$ of intrinsic and extrinsic AMR contributions as shown in the following.

### D. Intrinsic AMR component

Figure 3 allows us to tackle the major goal of this work: To determine the weight of scattering-based and scattering-independent components of AMR [see Eq. (3)]. For $\omega/2\pi >$ 2 THz, we find that the AMR contrast decreases by about 50% from 10 to 20 THz for both Ni$_{81}$Fe$_{19}$ and Ni$_{50}$Fe$_{50}$. The slope of this decrease is similar to that of the conductivity $\text{Re}\,\bar{\sigma}$. This observation and the discussion following Eq. (3) suggest that AMR contribution (A) is dominant for these films. In contrast, for Co, we find an AMR decrease of less than 10% from 10 to 20 THz, although the conductivity rolls off by more than 50% in this range. This finding and Eq. (3) indicate that the AMR of Co has a significant frequency-independent contribution (B).

To address this point quantitatively, we determined the weights $A$ and $B$ of the two AMR contributions (A) and (B) by fitting Eq. (3) to the measured $|\mathrm{AMR}(\omega)|$ (Fig. 3). Here, $A$, $B$ are the only fit parameters, whereas the value of the scattering time $\tau_\perp \approx \bar{\tau}$ is fixed by our analysis of the mean conductivity. Fitting was performed over both the full frequency range 0.2…28 THz and the high frequency range 8…28 THz (see Supplemental Note 1). With both procedures, we obtained excellent and consistent agreement of measured data and fits for all four investigated materials (Fig. 3).

The relevant parameters are summarized in Table 1. We find very small ratios $B/A$ of the order of $10^{-3}$ for $Ni_{81}Fe_{19}$, $Ni_{50}Fe_{50}$ and Ni. According to Eq. (3), an increase of $B/A$ would make the calculated curve $|\mathrm{AMR}(\omega)|$ even flatter at frequencies above 8 THz and result in less agreement with the experimental data [Fig. 3(a-c)]. Therefore, the intrinsic contribution (B) to the AMR of the $Ni_{81}Fe_{19}$, $Ni_{50}Fe_{50}$ and Ni samples is negligible.

We witness a strongly contrasting behavior for our Co thin film. A fit without the presence of an intrinsic contribution [$B = 0$ in Eq. (3)] yields a curve with significantly larger slope above 8 THz [grey dotted line in Fig. 3(d)], which agrees poorly with experimental data. A fit without this constraint results in very good agreement of Eq. (3) with the measured modulus of the AMR contrast for $B/A = 2.5 \pm 1.5$. Therefore, the intrinsic contribution to the AMR contrast [red dashed horizontal line in Fig. 3(d)] is a factor of about 2 larger than the extrinsic component. At the same time, Co exhibits a THz AMR of 0.7%, only 50% smaller than that of $Ni_{81}Fe_{19}$, which turns out to have the largest THz AMR of the four materials studied here. We have, thus, found direct experimental evidence for intrinsic contributions to AMR in a common ferromagnet.

## V. DISCUSSION

To summarize, we successfully measured AMR of thin films of the standard ferromagnets $Ni_{81}Fe_{19}$, $Ni_{50}Fe_{50}$, Ni and Co from DC to 28 THz. Our data can excellently be described by the Drude formula for the conductivity parallel and perpendicular to the sample magnetization. We identify two distinctly different contributions to AMR: (A) a frequency-dependent extrinsic component due to magnetization-dependent electron scattering and (B) a frequency-independent intrinsic component arising from magnetization-dependent electronic group velocities. While contribution (B) is usually neglected in Boltzmann-type models of AMR[2,3,5,12], it can be significant already at DC and even dominate the AMR above 20 THz in Co.

### A. Origin of the intrinsic AMR of Co

The question arises why contribution (B) to AMR is much larger in Co than in Ni, $Ni_{50}Fe_{50}$ and $Ni_{81}Fe_{19}$. We ascribe this distinctly different behavior to the crystal symmetry of the materials studied here. While crystalline Ni, $Ni_{50}Fe_{50}$ and $Ni_{81}Fe_{19}$ are cubic (fcc, point group m3m), Co has hexagonal symmetry (hcp, point group 6/mmm). The lower symmetry of Co allows for different values of observables for directions parallel and perpendicular to the c axis. Examples include the refractive index (making Co optically anisotropic already for $\boldsymbol{M} = 0$), the electron orbital angular momentum and spin-orbit coupling energies[52].

A strongly anisotropic spin-orbit coupling strength implies that the electronic band structure changes substantially when the magnetization $\boldsymbol{M}$ is parallel or perpendicular to the c axis. Therefore, the squared plasma frequency $\Omega_{\mathrm{pl}j}^2$, which is a summation of the squared electron band velocity component $j$ over the Fermi surface[34,35], should change strongly as well.

We put this expectation to test by numerically estimating the weight $B$ of the scattering-independent component (B) of the conductivity, that is, the $\boldsymbol{M}$-dependent variation of the squared plasma frequency $\Omega_{\mathrm{pl}j}^2$ of Co [see Eq. (3) and Appendix C]. Preliminary results indicate that when $\boldsymbol{M}$ is tilted out of the basal $x$-$y$ plane into a direction parallel to the c axis ($z$ axis) of Co, the plasma frequency $\Omega_{\mathrm{pl}z}$ decreases by a value of the order of 4%. In contrast, the calculations for Ni indicate that the plasma frequency varies significantly less than 1% as a function of the magnetization direction. Our numerical estimates, thus, confirm the expected variation of the plasma frequency $\Omega_{\text{pl}z}$ when $\boldsymbol{M}$ is rotated out of the basal plane of Co.

### B. Impact of polycrystallinity

We note that the samples of our experiment are polycrystalline. In a simplified picture, one can imagine this situation as an ensemble of three subsets of Co crystallites whose c axes point along either the $x$, $y$ or $z$ axis with the same probability of 1/3. For simplicity, we assume that only the magnetization component along the c axis will modify the conductivity. When the driving THz field $\boldsymbol{E}$ is applied along the z direction and the resulting current density $\boldsymbol{j}$ is measured along $\boldsymbol{E}$, the relevant conductivity $\sigma_{zz}$ changes only due to those crystallites whose c axis is parallel to the z axis. Therefore, the current density along $\boldsymbol{E}$ will change when $\boldsymbol{M}$ is rotated from $\boldsymbol{M} \parallel \boldsymbol{E}$ to $\boldsymbol{M} \perp \boldsymbol{E}$, and at least part of the AMR of the crystallites is inherited by the polycrystalline sample.

In a more rigorous way, the polycrystallinity of the sample can be taken into account by averaging the conductivity tensor over all crystal orientations while keeping the magnetization $\boldsymbol{M}$ fixed. One can, equivalently, perform a rotational average of the AMR tensor $G_{jklm} = (1/2)\partial^2 \sigma_{jk}/\partial M_l \partial M_m$ (see Appendix C). The elements $G_{jjll}$ are proportional to the change in $\Omega_{\text{pl}l}^2$ with respect to $M_l^2$. While the refractive index of polycrystalline Co in the absence of magnetic order ($\boldsymbol{M} = 0$) becomes completely isotropic, the AMR to a large extent survives the rotational averaging process. For polycrystalline Co, we estimate the scattering-independent AMR contrast by a linear combination of the numerically estimated tensor elements $G_{jjll}$. As detailed in Appendix C, we obtain an AMR contrast of $(0.8 \pm 0.5)\%$ for Co and 0% for Ni, which is in excellent agreement with the measured scattering-independent contribution of $B = (0.5 \pm 0.1)\%$ and $(0 \pm 0.07)\%$ (Table 1), respectively.

### C. Role of interband transitions

The Drude-Boltzmann theory of AMR and other electronic transport phenomena rely on intraband transitions: The probing THz field, possibly in conjunction with a phonon or an impurity, causes an electron to scatter from one Bloch state into another one in the same band of the electronic band structure. Above a certain probing frequency, however, interband transitions, that is, transitions between different bands, become operative.

Intraband transitions can often be well described by the Drude formula [Eq. (2)], and their contribution to the conductivity decays with $1/\omega$ for large enough frequencies. For interband transitions, we expect a different frequency dependence of the conductivity than for intraband transitions. Such crossover from intraband to vertical (i.e. wavevector-conserving) interband transitions was, for example, observed for the semimetal graphite already at frequencies between 10 and 20 THz (Ref. 53).

In our conductivity spectra (Fig. 3), however, we do not observe indications of interband transitions because we are able to well describe all measured curves by the simple Drude formula [Eq. (2)] over the full frequency range 0…28 THz. For Ni, this notion is consistent with earlier work[47] in which the onset of interband transitions was found at a photon energy of 0.15 eV (corresponding to 36 THz), which is outside the frequency range considered here. Similarly, for Co and Fe, previous studies report that the lowest interband transitions are at 0.18 eV (44 THz)[54,55] and 0.20 eV (48 THz)[48,49]. We conclude that in the materials studied here, intraband transitions dominate the response at least up to 30 THz. Therefore, our insights into intrinsic and extrinsic AMR contributions at THz frequencies can directly be transferred to the DC AMR.

### VI. CONCLUSIONS

In conclusion, low-noise broadband THz spectroscopy enables one to measure AMR from ~0.2 to tens of THz. The wide bandwidth provides access to important transport parameters. Our measurements reveal extrinsic and sizeable intrinsic contributions to the AMR contrast, thereby providing new and surprising insights into a mature effect.

Polycrystalline Co exhibits a sizeable intrinsic contribution which can consistently be ascribed to the crystalline anisotropy of the hexagonal (hcp) structure of Co crystallites. Our interpretation is supported by rotational averaging of the AMR tensor and numerical estimates. It highlights a strategy to identify materials with a large intrinsic AMR contribution, which is relevant for potential broadband THz spintronic applications.

Probing of the intrinsic AMR component is also highly interesting from a spectroscopic viewpoint because it reports on magnetic-order-induced variations of the electronic band structure. We anticipate that broadband THz AMR will be a highly useful, versatile and ultrafast probe of all flavors of magnetic order and transport parameters of spintronic materials. It can be applied to standard thin films, both crystalline[22] or polycrystalline, and under ambient conditions without the need of microstructuring and contacting. In particular, THz AMR should also be applicable to metallic antiferromagnets such as CuMnAs and $Mn_2Au$ that have recently moved into the focus of spintronics research[11,56].

As our THz radiation is pulsed, THz AMR can be measured with a time resolution down to 100 fs. This feature opens up the exciting possibility to monitor material-relevant parameters on the natural time scales of spin, electron and lattice dynamics[57]. In this way, THz AMR complements other recently developed ultrafast spintronic techniques such as THz anomalous Hall effect[50,58,59], THz tunnel magnetoresistance[25], THz giant magnetoresistance[33] and magnetization-dependent THz emission[43,60,61,62,63] that have provided new insights into the dynamics of spin transport and spin-to-charge-current conversion. Finally, because the THz range coincides with a variety of excitations (such as phonons and magnons), the method presented here allows us to study the impact of such resonances on magneto-transport at their natural frequencies.

## ACKNOWLEDGMENTS

We acknowledge funding by the German Research Foundation through collaborative research centers SFB TRR 227 "Ultrafast spin dynamics" (projects A05, B02 and B04) and SFB TRR 173 "Spin+X" (projects A01 and B02). We thank the ERC for support through the Horizon2020 projects CoG TERAMAG/Grant No. 681917, SyG 3D MAGiC/Grant No. 856538 and ASPIN/Grant No. 766566. We acknowledge financial support from the Horizon 2020 Framework Programme of the European Commission under FET-Open Grant No. 863155 (s-Nebula).

## APPENDIX A: EXPERIMENTAL DETAILS

### 1. Sample growth and characterization

All samples were deposited by sputtering or thermal evaporation techniques[64] on thermally oxidized Si|$SiO_2$(100 nm) substrates. Details on growth and characterization can be found in Supplemental Note 3. In brief, the $Ni_{81}Fe_{19}$ thin film (thickness of 8 nm) was grown by DC sputtering (sputter power of 800 W, Ar pressure of 0.5 Pa). X-ray diffraction ($\theta$-$2\theta$ scans) reveals a very weak (111) reflection, indicating a crystallite size of about 3 nm. A weak (220) reflection confirms the polycrystalline growth of the sample. The layer thicknesses were inferred from the X-ray reflectometry measurements.

The $Ni_{50}Fe_{50}$(10 nm) layer was deposited by DC magnetron sputtering (sputter power 30 W, Ar pressure 0.4 Pa). After the sputtering process, the sample was capped with an MgO(7 nm) layer grown by in-situ molecular beam epitaxy (MBE) and electron-beam evaporation and by an $Al_2O_3$(5 nm) layer by ex-situ atomic layer deposition. The Co(10 nm) film was grown by thermal evaporation in ultrahigh vacuum (MBE) and capped by MgO(5 nm) and Al$_2$O$_3$(5 nm) using the techniques described above. The Ni(10 nm) film was prepared by thermal evaporation in a vacuum chamber and capped by Al(3 nm), which fully oxidizes under ambient conditions[65]. The crystal structure of the Co film was monitored during growth using reflection high-energy electron diffraction. We find a polycrystalline hcp structure with random crystal orientation.

## 2. THz conductivity measurements

Our sample system is a stack S|F|A consisting of a metal thin film F (thickness $d$) between substrate S (refractive index $n_S$) and air A (refractive index $n_A$). To determine the conductance of F, we conduct transmission measurements [see Fig. 1(b)]. In a first measurement, we characterize the THz field $\boldsymbol{E}$ directly behind the F layer. As the field $\boldsymbol{E}_{\text{inc}}$ incident on the sample is unknown, we conduct a second measurement on a reference sample S|R|A where the sample film F is replaced by a reference film R with known refractive index. In practice, the reference measurement is performed in sample regions where no metal film is deposited. Thus, our reference material is air (R = A), whose refractive index equals 1 to very good approximation.

In our setup, the field $\boldsymbol{E}_{\text{inc}} = E_{\text{inc}}\boldsymbol{u}$ is normally incident onto the sample and linearly polarized parallel to the vertical unit vector $\boldsymbol{u}$ [Fig. 1(b)]. The field $\boldsymbol{E}$ behind the sample is projected onto the same direction $\boldsymbol{u}$ by means of a polarizer. We, thus measure a signal $S$ which is related to the projection $\boldsymbol{u} \cdot \boldsymbol{E}$ through the transfer function of our setup. Likewise, the signal $S_{\text{ref}}$ from the reference sample is obtained. By dividing the signals $S(\omega)$ and $S_{\text{ref}}(\omega)$ in the frequency domain, the setup transfer function cancels. As derived in Appendix B, the ratio $S(\omega)/S_{\text{ref}}(\omega)$ is related to sample-intrinsic parameters by

$$\frac{S(\omega)}{S_{\text{ref}}(\omega)} = \frac{1}{1 + Z_{\text{ref}}(\omega)\overline{\sigma}(\omega)d} \left[1 + \frac{\text{AMR}(\omega)\cos^2\alpha}{1 + [Z_{\text{ref}}(\omega)\overline{\sigma}(\omega)d]^{-1}}\right] \quad (A1)$$

where $Z_{\text{ref}} = Z_0/[n_S(\omega) + n_A(\omega)]$ is the impedance of the reference sample.

As the AMR contrast $\text{AMR}(\omega)$ amounts to only a few per cent, the second term in the square bracket of Eq. (A1) is much smaller than the first one. Consequently, we determine the first term by a simple transmission measurement through the sample averaged over all magnetization directions and through the reference without metal film. We obtain the familiar Tinkham formula[46]

$$\frac{\overline{S}(\omega)}{S_{\text{ref}}(\omega)} = \frac{1}{1 + Z_{\text{ref}}(\omega)\overline{\sigma}(\omega)d} \quad (A2)$$

which implies that

$$\overline{\sigma}(\omega) = \frac{1}{Z_{\text{ref}}(\omega)d}\left(\frac{S_{\text{ref}}(\omega)}{\overline{S}(\omega)} - 1\right). \quad (A3)$$

To measure the AMR-related term in Eq. (A1), we modulate $\alpha$ between 0 and 90°. We obtain $\Delta S(\omega) = S(\omega, 0°) - S(\omega, 90°)$ and, thus,

$$\frac{\Delta S(\omega)}{\overline{S}(\omega)} = \frac{\text{AMR}(\omega)}{1 + [Z_{\text{ref}}(\omega)\overline{\sigma}(\omega)d]^{-1}}, \quad (A4)$$

which is equivalent to Eq. (6) of the main text.

## 3. Error considerations

**THz measurements.** The uncertainties of the fit parameters $\overline{\sigma}(0)$, $\overline{\tau}$ [Eq. (2)] and $A$, $B$ [Eq. (3)] are given by the uncertainties of the signals $\Delta S$ and $\overline{S}$ and the statistics of the fit procedure. They are summarized in Table 1 and Supplemental Table S1.

The precision of $\overline{S}$ is estimated by the standard error of repeated measurements of this signal. To estimate the uncertainties of $\Delta S$, two contributions were considered. The first one is the statistical error of $\Delta S$ that arises from the shot noise of our measurement. It is estimated by the constant noise floor outside the signal bandwidth of the THz-emitter-detector configuration used. An example of the noise floor for $Ni_{81}Fe_{19}$ is shown in Supplemental Fig. S5.

The second error contribution to $\Delta S$ arises from the finite precision with which the magnets for rapid modulation of the angle $\alpha$ of the external magnetic field could be positioned. As a consequence, the minimum angle $\alpha_{\min}$ and the maximum angle $\alpha_{\max}$ deviated from the target angles 0° and 90°, respectively. This systematic error only results in an overall rescaling of $\Delta S$ by an estimated upper limit of 30%. Importantly, it does not affect the frequency dependence of $\Delta S$. It may, however, differ between the measurements in the ranges 0.2…2 THz and 8…28 THz where the permanent magnet and the electromagnet were repositioned. This issue was tackled by fitting Eq. (3) over the full frequency range 0.2…28 THz and the higher range 8…28 THz only. We obtained consistent results, as summarized in the Supplemental Note 1 and Supplemental Table S1.

**DC measurements.** For the electrical measurements of $\sigma_{DC}$ by the van-der-Pauw method, the measurement error is of the order of 5% and predominantly arises from the nonvanishing size of the contacts and their positioning within the sample perimeter[40,41,42]. The error of the electrical measurement of the AMR contrast by our four-point approach is governed by the uncertainty of the direction of the current flow between electrical contacts, the directional homogeneity of the external magnetic field of the Halbach array and the fit statistics of the raw data (see Fig. S1).

**Comparison DC THz.** From Fig. 3, we observe that the values of the THz AMR contrast below 2 THz are smaller than the DC AMR contrast for $Ni_{50}Fe_{50}$, Ni and Co. This behavior can be explained by a deviation of $\alpha_{\max} - \alpha_{\min}$ from 90°, which leads to a reduction of the measured THz AMR contrast. For $Ni_{81}Fe_{19}$, we observe the opposite behavior which, we believe, arises from the different aspect ratios of the rectangular samples used for the DC AMR measurements. While the aspect ratio of the $Ni_{81}Fe_{19}$ sample was close to 1:1, it was roughly 4:1 for the other samples. As a consequence, the current flow in the $Ni_{81}Fe_{19}$ sample was less homogeneous, resulting in an apparently smaller measured DC AMR contrast.

## APPENDIX B: DERIVATION OF EQS. (3) AND (A1)

The following derivations refer to complex-valued quantities in the frequency domain. For the sake of simplicity, the argument $\omega$ is omitted.

### 1. Derivation of Eq. (3)

We first rewrite the Drude formula[30,31] [Eq. (2)] as

$$\sigma_j(\omega) = \frac{\Omega_{\mathrm{pl}j}^2/Z_0 c}{\tau_j^{-1} - i\omega} =: \frac{N_j}{D_j} \tag{B1}$$

where $j = 1, 2$ refers to the $\perp$ and $\parallel$ configuration, respectively. Linearization with respect to $\Delta(\Omega_{\mathrm{pl}}^2) = \Omega_{\mathrm{pl}2}^2 - \Omega_{\mathrm{pl}1}^2 \ll \Omega_{\mathrm{pl}j}^2$ and $\Delta\tau = \tau_2 - \tau_1 \ll \tau_j$ yields

$$-\mathrm{AMR} = \frac{\Delta\sigma}{\sigma_1} = \frac{\sigma_2 - \sigma_1}{\sigma_1} = \frac{\Delta N}{N_1} - \frac{\Delta D}{D_1} = \frac{\Delta(\Omega_{\mathrm{pl}}^2)}{\Omega_{\mathrm{pl}1}^2} + \frac{\Delta\tau}{\tau_1^2} \frac{1}{\tau_1^{-1} - i\omega} = -B - \frac{A}{1 - i\omega\tau_1} \tag{B2}$$

with $B = -\Delta(\Omega_{\text{pl}}^2)/\Omega_{\text{pl}1}^2$ and $A = -\Delta\tau/\tau_1$. Further analysis shows that the error of the linearization is of the order of $\Delta(\Omega_{\text{pl}}^2)\Delta\tau$, which is negligible here.

## 2. Derivation of Eq. (A1)

**Wave equation.** In our setup, the incident THz pulse propagates along the $z$ axis, which is perpendicular to the sample plane [see Fig. 1(b)]. Therefore, $z$ is the only relevant spatial coordinate, and we choose its origin such that the metal film F is located between $z = 0$ and $d$. We assume that substrate S and air A are optically isotropic and homogeneous and can, thus, be described by scalar refractive indices $n_S$ and $n_A$, respectively. The metal thin film F, in contrast, is allowed to be inhomogeneous along $z$ and optically anisotropic. It is adequately described by the conductivity tensor (matrix) $\underline{\sigma}(z)$.

In frequency space, the THz field $\boldsymbol{E}(z)$ is determined by the wave equation[66]

$$\left(\partial_z^2 + \underline{\beta}^2\right)\boldsymbol{E} = \boldsymbol{Q}_{\text{ext}}. \tag{B3}$$

Here, $\boldsymbol{Q}_{\text{ext}}$ quantifies the sample-external source of the incident THz wave, and the squared wavenumber matrix $\underline{\beta}^2(z)$ captures the linear-optical properties of the system. Its difference to the reference system fulfills

$$\left(\underline{\beta}^2 - \beta_{\text{ref}}^2\right)(z) = \frac{iZ_0}{\omega/c}\underline{\sigma}(z) \tag{B4}$$

where $Z_0 \approx 377\,\Omega$ is the free-space impedance. The reference system is the sample without metal film, that is, just the substrate and air half-spaces. We rewrite Eq. (B3) as

$$\left(\partial_z^2 + \beta_{\text{ref}}^2\right)\boldsymbol{E} = \boldsymbol{Q}_{\text{ext}} + \boldsymbol{Q} := \boldsymbol{Q}_{\text{ext}} + \frac{Z_0}{i\omega/c}\underline{\sigma}\boldsymbol{E} \tag{B5}$$

where the term $\boldsymbol{Q}(z)$ quantifies the source of the field component that arises from the response of the metal film with conductivity $\underline{\sigma}$. By inverting the operator $\partial_z^2 + \beta_{\text{ref}}^2$ in Eq. (B5), one obtains the integral equation[66]

$$\boldsymbol{E}(z) = \boldsymbol{E}_{\text{ref}}(z) + \int dz'\, \underline{\mathcal{G}}_{\text{ref}}(z, z')\,\boldsymbol{Q}(z') \tag{B6}$$

where $\underline{\mathcal{G}}_{\text{ref}}(z, z')$ is the optical Green's function of the reference sample. Equation (B6) has a clear physical interpretation: The total THz field $\boldsymbol{E}(z)$ is the sum of the field $\boldsymbol{E}_{\text{ref}}(z)$ of the reference sample (no metal film) plus the field generated by the field-induced currents in the metal.

**Thin-film approximation.** To solve Eq. (B6), we apply the so-called thin-film approximation and assume that the field is constant throughout the thickness of the metal film, that is,

$$\boldsymbol{E}(z) = \boldsymbol{E}, \qquad \boldsymbol{E}_{\text{ref}}(z) = \boldsymbol{E}_{\text{ref}} \tag{B7}$$

in the vicinity of $z = 0$. This assumption is fulfilled if the thickness of the metal film is much smaller than the wavelength and the attenuation length of the THz wave inside the metal. Likewise, for $z \approx 0$, the Green's function of the reference sample becomes a $(z, z')$-independent scalar, that is,

$$\underline{\mathcal{G}}_{\text{ref}}(z, z') = g_{\text{ref}} := \frac{1}{i \cdot (\beta_S + \beta_A)} \tag{B8}$$

where $\beta_j = n_j\omega/c$ is the wavenumber of the substrate ($j = S$) and air ($j = A$). By combining Eqs. (B5), (B7) and (B8) with Eq. (B6), we finally obtain the total field

$$\boldsymbol{E} = \frac{1}{1 + Z_{\text{ref}}\underline{G}_F}\boldsymbol{E}_{\text{ref}}, \tag{B9}$$

where

$$Z_{\text{ref}} = \frac{Z_0}{n_S + n_A} \quad \text{and} \quad \underline{G}_F = \int dz' \, \underline{\sigma}(z'), \tag{B10}$$

respectively, is the impedance of the reference system close to the S/A interface and the (anisotropic) conductance of the metal film F. For a homogeneous film with $z$-independent conductivity $\underline{\sigma}$, we have $\underline{G}_F = \underline{\sigma}d$.

**Application to our sample.** For our magnetic thin film F, the conductivity can be split according to $\underline{\sigma} = \sigma_0 + \Delta\underline{\sigma}$ where $\sigma_0$ is the isotropic conductivity in the absence of magnetization ($\mathbf{M} = 0$), while the anisotropic part $\Delta\underline{\sigma}$ captures all magnetoresistive effects. By linearizing Eq. (B9) with respect to $\Delta\underline{\sigma}$, we find

$$\mathbf{E} = \frac{\mathbf{E}_{\text{ref}}}{1 + Z_{\text{ref}} G_{F0}} - \frac{Z_{\text{ref}} \Delta \underline{G}_F \mathbf{E}_{\text{ref}}}{(1 + Z_{\text{ref}} G_{F0})^2}, \tag{B11}$$

where $G_{F0} = \int dz' \, \sigma_0(z')$ is the conductance for $\mathbf{M} = 0$, and $\Delta\underline{G}_F = \int dz' \, \Delta\underline{\sigma}(z')$ is the magnetoresistive contribution.

In our experiment, the incident field and, thus, reference are linearly polarized parallel to the $x$ axis [see Fig. 1(b)]. Therefore, they can be written as $\mathbf{E}_{\text{inc}} = E_{\text{inc}}\mathbf{u}$ and $\mathbf{E}_{\text{ref}} = E_{\text{ref}}\mathbf{u}$ where $\mathbf{u}$ is the unit vector ($|\mathbf{u}|^2 = \mathbf{u}^2 = 1$) of the $x$ axis. In addition, the field $\mathbf{E}$ behind the sample is projected onto the same direction $\mathbf{u}$ by a polarizer, resulting in $E = \mathbf{u} \cdot \mathbf{E}$. We multiply Eq. (B11) with $\mathbf{u}$ from the left side and arrive at

$$\frac{E}{E_{\text{ref}}} = \frac{1}{1 + Z_{\text{ref}} G_{F0}} - \frac{Z_{\text{ref}} \Delta G_{F\mathbf{uu}}}{(1 + Z_{\text{ref}} G_{F0})^2}, \tag{B12}$$

where $\Delta G_{F\mathbf{uu}} = \mathbf{u} \cdot \Delta\underline{G}_F \mathbf{u}$ is the magnetization-induced change in the conductance projected onto $\mathbf{u}$.

For our polycrystalline, homogeneous and ferromagnetic F layer, we have $G_{F0} = \sigma_0 d$ and $\Delta\underline{G}_F = \Delta\underline{\sigma}d$. According to Eq. (C2), the magnetoresistive part $\Delta\underline{\sigma}$ of the conductivity tensor fulfills

$$\Delta\underline{\sigma}\mathbf{u} = a\mathbf{M} \times \mathbf{u} + b\mathbf{M}(\mathbf{M} \cdot \mathbf{u}). \tag{B13}$$

The first term is the anomalous Hall effect[50] with $\mathbb{I}$ denoting the unity matrix, and the second term is the AMR. The contribution of isotropic magnetoresistance can be added to $\sigma_0$, but was neglected here because it is much smaller than $\sigma_0$. The constants $a$ and $b$ are material-specific, and $b$ is directly related to the AMR tensor.

Equations (B13) and (C3) imply that $\mathbf{u} \cdot \Delta\underline{\sigma}\mathbf{u} = b\mathbf{M}^2 \cos^2 \alpha = (\sigma_\parallel - \sigma_\perp) \cos^2 \alpha$ where $\alpha$ is the angle between $\mathbf{M}$ and $\mathbf{u}$ [Fig. 1(b)]. Therefore, projection of the transmitted field onto $\mathbf{u}$ does not contain any contribution of the anomalous Hall effect since these field changes are perpendicular to $\mathbf{u}$. By substituting this result into Eq. (B12), we obtain

$$\frac{E}{E_{\text{ref}}} = \frac{1}{1 + Z_{\text{ref}}\bar{\sigma}d} - \frac{Z_{\text{ref}} d(\sigma_\parallel - \sigma_\perp) \cos^2 \alpha}{(1 + Z_{\text{ref}}\bar{\sigma}d)^2}. \tag{B14}$$

Here, we replaced $\sigma_0$ by $\bar{\sigma} = (\sigma_\parallel + \sigma_\perp)/2$ in the denominators of Eq. (B14) with negligible error because $|\sigma_\parallel - \sigma_\perp| \ll \sigma_0 \approx \bar{\sigma}$. Along with the AMR contrast $\text{AMR} = -(\sigma_\parallel - \sigma_\perp)/\bar{\sigma}$, Eq. (B14) turns into

$$\frac{E(\omega)}{E_{\text{ref}}(\omega)} = \frac{1}{1 + Z_{\text{ref}}\bar{\sigma}d}\left[1 + \frac{\text{AMR}\cos^2 \alpha}{1 + (Z_{\text{ref}}\bar{\sigma}d)^{-1}}\right], \tag{B15}$$

which is the desired relationship of the main text.

# APPENDIX C: AMR OF POLYCRYSTALLINE SAMPLES

## 1. Symmetry analysis

The AMR tensor of an arbitrary ferromagnetic material is defined by $G_{jklm} = (1/2)\partial^2 \sigma_{jk}/\partial M_l \partial M_m$. Depending on the point symmetry group of the material, a substantial number of tensor elements is strictly zero or depends on each other, thereby resulting in a relatively small number of independent tensor elements. For the hexagonal crystal structure of Co (hcp, point group 6/mmm), we have[67,68]

$$\begin{pmatrix}\sigma_{xx}\\ \sigma_{yy}\\ \sigma_{zz}\\ \sigma_{yz}\\ \sigma_{zx}\\ \sigma_{xy}\end{pmatrix} = \begin{pmatrix} G_{xxxx} & G_{xxyy} & G_{xxzz} & 0 & 0 & 0 \\ G_{xxyy} & G_{xxxx} & G_{xxzz} & 0 & 0 & 0 \\ G_{zzxx} & G_{zzxx} & G_{zzzz} & 0 & 0 & 0 \\ 0 & 0 & 0 & 2G_{yzyz} & 0 & 0 \\ 0 & 0 & 0 & 0 & 2G_{yzyz} & 0 \\ 0 & 0 & 0 & 0 & 0 & G_{xxxx}-G_{xxyy} \end{pmatrix} \begin{pmatrix} M_x^2 \\ M_y^2 \\ M_z^2 \\ M_y M_z \\ M_z M_x \\ M_x M_y \end{pmatrix}, \quad (C1)$$

where the $z$ axis is oriented along the c axis of Co.

For the cubic crystal structure of Ni, Ni$_{50}$Fe$_{50}$ and Ni$_{81}$Fe$_{19}$ (fcc, point group m3m), one has the additional constraints $G_{xxyy} = G_{xxzz} = G_{zzxx}$, $G_{xxxx} = G_{zzzz}$ and $G_{xxxx} - G_{xxyy} = 2G_{yzyz}$. Therefore, two independent elements such as $G_{xxyy}$ and $G_{yzyz} = G_{zxzx} = G_{xyxy}$ completely determine the AMR tensor in this case.

## 2. Rotational averaging

The AMR tensor of a polycrystalline material is obtained by rotational averaging[69,70] of the tensor $G_{jklm}$ of the crystalline material. The resulting tensor $\langle G \rangle_{jklm}$ fulfills the same symmetry constraints as the AMR tensor of a cubic crystal with point group m3m. Thus, knowledge of the two independent elements $\langle G \rangle_{xyxy}$ and $\langle G \rangle_{xxyy}$ is sufficient. For a ferromagnetic material of this symmetry class, the current density $\bm{j}$ induced by an electric field $\bm{E}$ can up to second order in the magnetization $\bm{M}$ compactly be written as[67,68]

$$\bm{j} = \underline{\sigma}\bm{E} = \sigma_0 \bm{E} + a \bm{M} \times \bm{E} + b\bm{M}(\bm{M}\cdot\bm{E}) + c\bm{M}^2 \bm{E}. \quad (C2)$$

Here, the first term on the right-hand side is the current in the absence of magnetic order, the second term with constant $a$ is the anomalous Hall effect, the third term is the AMR with $b = 2\langle G \rangle_{xyxy}$, and the last term is an isotropic magnetoresistance with $c = \langle G \rangle_{xxyy}$. For a thin film of this material with $\bm{M}$ and $\bm{E}$ in the film plane, the conductivity is $\sigma_\perp = \sigma_0 + c\bm{M}^2$ if $\bm{M} \perp \bm{E}$, whereas it is $\sigma_\parallel = \sigma_\perp + b\bm{M}^2$ if $\bm{M} \parallel \bm{E}$. We, thus, have

$$\sigma_\parallel - \sigma_\perp = b\bm{M}^2 = 2\langle G \rangle_{xyxy}\bm{M}^2. \quad (C3)$$

For polycrystalline Co, we performed rotational averaging[69,70] of the AMR tensor and obtained

$$30\langle G \rangle_{xyxy} = 7G_{xxxx} + 2G_{zzzz} - 5G_{xxyy} - 2G_{xxzz} - 2G_{zzxx} + 12G_{yzyz} \quad (C4)$$

whereas for polycrystalline Ni, we found

$$\langle G \rangle_{xyxy} = G_{xyxy}. \quad (C5)$$

## 3. AMR estimate of polycrystalline samples

We conducted *ab initio* calculations of the plasma frequency of crystalline Ni and Co as detailed in Supplemental Note 2. These values can be compared to the weight $B$ of the intrinsic AMR contribution we measured on polycrystalline samples of Ni and Co. For Ni, the calculated plasma frequencies are independent of the magnetization direction (see Supplemental Table S3), consistent with $B = (0 \pm 0.07)\%$ as inferred from our measurements.

To deal with Co, we assume that its AMR exclusively arises from intrinsic contributions and use Eqs. (1), (C3) and (C4) to write

$$-B = \frac{\Delta\sigma}{\bar{\sigma}} = \frac{2}{\bar{\sigma}} \langle G \rangle_{xyxy} \boldsymbol{M}^2 = \frac{2}{\bar{\sigma}} \sum \boldsymbol{M}^2 G_{ijkl} a_{ijkl} \approx \frac{2}{\bar{\sigma}} \sum \boldsymbol{M}^2 G_{zzjj} a_{zzjj}. \tag{C6}$$

The coefficients $a_{ijkl}$ are related to the rotational averaging [see Eq. (C4)], and in the last step of Eq. (C6), we neglected $G_{yzyz}$. According to our *ab initio* calculations (see Supplemental Table S3), we infer that only the plasma frequencies and, thus, the conductivities $\sigma_{zz}$ [see Eq. (2)] along the c axis ($z$ axis) of crystalline Co depend on the direction of the magnetization $\boldsymbol{M}$. On the other hand, from Eq. (C4), we see that $a_{zzxx} = -1/15$, $a_{zzyy} = 0$ and $a_{zzzz} = 1/15$. With these conditions, Eq. (C6) becomes

$$-B \approx \frac{2}{15\bar{\sigma}} (\boldsymbol{M}^2 G_{zzzz} - \boldsymbol{M}^2 G_{zzxx}) = \frac{2}{15} \frac{\sigma_{zz;zz} - \sigma_{zz;xx}}{\bar{\sigma}}. \tag{C7}$$

In the last step, we used that $\boldsymbol{M}^2 G_{zzjj} = \sigma_{zz;jj} - \sigma_{zz}^0$ where $\sigma_{zz;jj}$ is the z conductivity when the sample is magnetized along the $j$ direction, and $\sigma_{zz}^0$ is the z conductivity for $\boldsymbol{M} = 0$. We finally assume that the current relaxation time of Co is isotropic for $\boldsymbol{M} = 0$. By combining Eqs. (2) and (C7) and using Supplemental Table S3, we obtain

$$B \approx -\frac{2}{15} \frac{\Omega_{\text{pl}zz;zz}^2 - \Omega_{\text{pl}zz;xx}^2}{\Omega_{\text{pl}}^2} = (0.8 \pm 0.5)\%, \tag{C8}$$

which is in excellent agreement with the experimentally determined value of $B = (0.5 \pm 0.1)\%$.